\documentclass{epl}

\usepackage{graphicx}
\usepackage{dcolumn}
\usepackage{bm}
\input{epsf}

\title{Persistence of small-scale anisotropy of magnetic turbulence as observed in the solar wind}

\author{Luca Sorriso-Valvo$^{1,2}$, Vincenzo Carbone$^2$, Roberto Bruno$^3$, and Pierluigi Veltri$^2$}
\institute{
\inst{1} Liquid Crystal Laboratory - INFM/CNR, Ponte P. Bucci, Cubo 31C, 87036 Rende (CS), Italy \\
\inst{2} Dipartimento di Fisica - Univerist\`a della Calabria, and Istituto Nazionale di Fisica della Materia, Unit\`a di Cosenza, Ponte P. Bucci, Cubo 31C, 87036 Rende (CS), Italy \\
\inst{3} Istituto di Fisica dello Spazio Interplanetario- INAF, Via Fosso del Cavaliere, 00133 Roma, Italy.
}

\begin{document}
\maketitle

\begin{abstract}

The anisotropy  of magnetophydrodynamic turbulence is investigated by using solar wind data from the Helios 2 spacecraft. We investigate the behaviour of the complete high-order moment tensors of magnetic field increments and we compare the usual longitudinal structure functions which have both isotropic and anisotropic contributions, to the fully anisotropic contribution. Scaling exponents have been extracted by an interpolation scaling function. Unlike the usual turbulence in fluid flows, small-scale magnetic fluctuations remain anisotropic. We discuss the radial dependence of both anisotropy and intermittency and their relationship.

\end{abstract}

The idea that small scales were statistically isotropic to a high degree, in spite of the anisotropy of large scales, is at the heart of the concept of universality underlying scaling behaviour in turbulent fluid flow \cite{k41}. As experiments became increasingly sophisticated, evidence for the presence of anisotropy for both scalar fields \cite{sreen} and velocity fields \cite{shen} has grown up. To what extent scaling laws extracted from experiments are approximated by the scaling laws obtained from (isotropic) cascade models~\cite{frisch}, became an urgent issue. A workable benchmark is represented by the following conjecture \cite{kurien1,kurien2}. Suppose that for a given statistical measurable quantity, scaling features can be splitted into an isotropic and an anisotropic part. We might ask to what extent the ratio between the anisotropic and the isotropic part vanishes as the scale goes down. Of course if we are able to define quantities for which the isotropic part is exactly zero, we can measure scaling laws related entirely to the anisotropic part. A formal analysis to reveal the isotropic and anisotropic contributions to structure functions can be performed by considering their SO(3) decomposition~\cite{so3}. This is based in a foliation of various $j$-sectors of the linear space generated by the structure functions. For a recent review on anisotropy in turbulent flows see~\cite{biferale}.

Anisotropy of turbulence in charged fluids, described by Magnetohydrodynamic (MHD) \cite{biskamp,CV1,anisoB}, can be due to a background magnetic field $\mathbf{B}_0$, or a magnetic field at large scales. Since $\mathbf{B}_0$ cannot be cancelled out by a Galilean transformation, its direction becomes privileged with respect to other directions. Then MHD turbulence is anisotropic to a very large extent. 
A simple shell model for anisotropic MHD~\cite{CV1} showed that even if the energy injection rate is isotropic, or isotropy is restored at large scales, intermediate scales and mainly small scales are strongly anisotropic and the return-to-isotropy at small scales does not hold in MHD.
This important result has been confirmed both by numerical simulations~\cite{anisoB}, and more recently by analysis of Ulysses spacecraft in fast solar wind~\cite{bigazzi}, where it is shown that strongly intermittent and anisotropic events persist even at high frequencies.

The study of the nature of the low-frequency magnetic fluctuations in the interplanetary space, in the frequency range $10^{-7}$ Hz $\leq f \leq 1$ Hz, is important because it represents the only way to understand the behaviour of MHD turbulence in a frequency domain which is not accessible to any laboratory on Earth \cite{RevMT,RevBC}. Anisotropy of MHD fluctuations in the solar wind, due to the background magnetic field transported away from the sun by the plasma flow \cite{aniso,bavassa82}, has been investigated through the minimum variance analysis \cite{minimumvariance}. This method gives informations about the energy distribution among the different components of magnetic fluctuations. It is based on the determination of eigenvalues and eigenvectors of the variance matrix $C_{\alpha\beta} = \langle B_\alpha B_\beta \rangle$ (brackets mean ensemble averages). During the so-called Alfv\'enic periods \cite{RevBC} one of eigenvalues is much smaller than the other two $\lambda_3 \ll \lambda_2 < \lambda_1$, for example the ratios are $\lambda_1:\lambda_2:\lambda_3 = 10:3:1$ \cite{bavassa82}. The minimum variance direction turns out to be almost parallel to $\mathbf{B}_0$, so the magnetic fluctuations lie in a plane approximately perpendicular to $\mathbf{B}_0$. These old studies have been used to investigate the spectral properties of the second-order tensor components, and to investigate the radial evolution of anisotropy. Here we investigate the scaling properties of anisotropy of the interplanetary magnetic field, by using the data analysis technique described extensively in Ref.s \cite{kurien1,kurien2} and \cite{shen}. This technique allows us to define the scaling properties of anisotropy of the interplanetary space fluctuations, as far as scaling exponents of high-order tensors are considered, so that intermittency is considered.

The data we use in this work are $6$~second measurements of the Interplanetary Magnetic Field (IMF), as recorded by Helios~2 flux-gate magnetometer. Velocity measurements from Helios~2 instruments are only available at lower resolution so that the statistical accuracy does not allow to perform the analysis presented in this work. The spacecraft followed an ecliptic orbit with perihelion and aphelion located at about 0.3 and 1 AU, respectively. The peripheral speed of the spacecraft along its orbit was much less than the solar wind speed so that the latter can be still be considered aligned to the radial direction from the sun. Helios 2 magnetic field measurements are given in a spacecraft centered reference system whose $x$ axis is along the radial direction and positive towards the sun, the $y$ axis lies on the ecliptic plane and is positive in the opposite direction with respect to the orbital motion, and the $z$ axis is perpendicular to the ecliptic and completes the right-handed reference system. In this reference system, field differences at different lags $r$ along the $x$ direction can be obtained as $\Delta B_\alpha(r) = B_\alpha(x+r) - B_\alpha(x)$. The period of time used in this analysis spans across the first 4 months of 1976 when the spacecraft moved from 1 to 0.3 AU. During this period, for which the solar activity was near the minimum, the solar wind was clearly characterized by alternating fast streams and slow wind (cfr. e.g. Ref.s \cite{RevMT,RevBC}). The different turbulent state of the fluctuations observed within these samples of fast and slow wind and the strong radial evolution experienced by these fluctuations (cfr. e.g. Ref.s \cite{RevMT,RevBC}) convinced us about the necessity to analyze these data separately per wind speed and radial distance \cite{sorriso99}, avoiding also to include velocity shear interfaces and the only shock that was present in this particular data set. As a matter of fact, the radial evolution is crucial for anisotropy studies, both because of the natural evolution of intermittency \cite{bruno_rad,bruno_aniso}, and because of the rotation of the mean field along the Parker spiral. Thus, two different samples for each wind type have been selected, at two well separated distances from the Sun, namely at $0.9$ AU and $0.3$ AU ($1$ AU $\simeq 1.5\times 10^8$ km), each data sample including about three days of records ($\sim 4.3 \times 10^4$ data points). The switch between space and time sampling is obtained from the Taylor hypothesis, so that $x = -t \langle v_x \rangle$, the average speed being computed for each stream.

In order to measure the contribution from the  anisotropic part of the field differences, the whole independent elements of the $n$-th order tensor, computed from time series, must be studied
\begin{equation}
S_{\alpha_1,\alpha_2,\cdots,\alpha_n}(r) = \langle \left[B_{\alpha_1}(x+r) - B_{\alpha_1}(x) \right]
\left[B_{\alpha_2}(x+r) - B_{\alpha_2}(x) \right] \cdots \left[B_{\alpha_n}(x+r) - B_{\alpha_n}(x) \right]\rangle
\label{Sn}
\end{equation}
For example, when $n=2$ and $\alpha_1=\alpha_2=x$ the usual longitudinal second-order structure function is recovered. The latter have been studied in the past, together with the longitudinal higher order moments \cite{RevBC,bruno_rad}. For each scale, some of the components of the tensor can be proven to have contribution from the anisotropic part of the field only (we will use in this work $S_{xz}$ for the second order, $S_{xxxz}$ and $S_{xzzz}$ for the fourth order structure functions)\cite{kurien1,kurien2,bigazzi}. It is thus possible to compare the scaling behavior of such components with those of the longitudinal structure functions (namely $S_{xx}$ and $S_{xxxx}$), which includes both isotropic and anisotropic contributions \cite{kurien1,kurien2,bigazzi}. 

\begin{figure}
\epsfxsize=14cm
\centerline{\epsffile{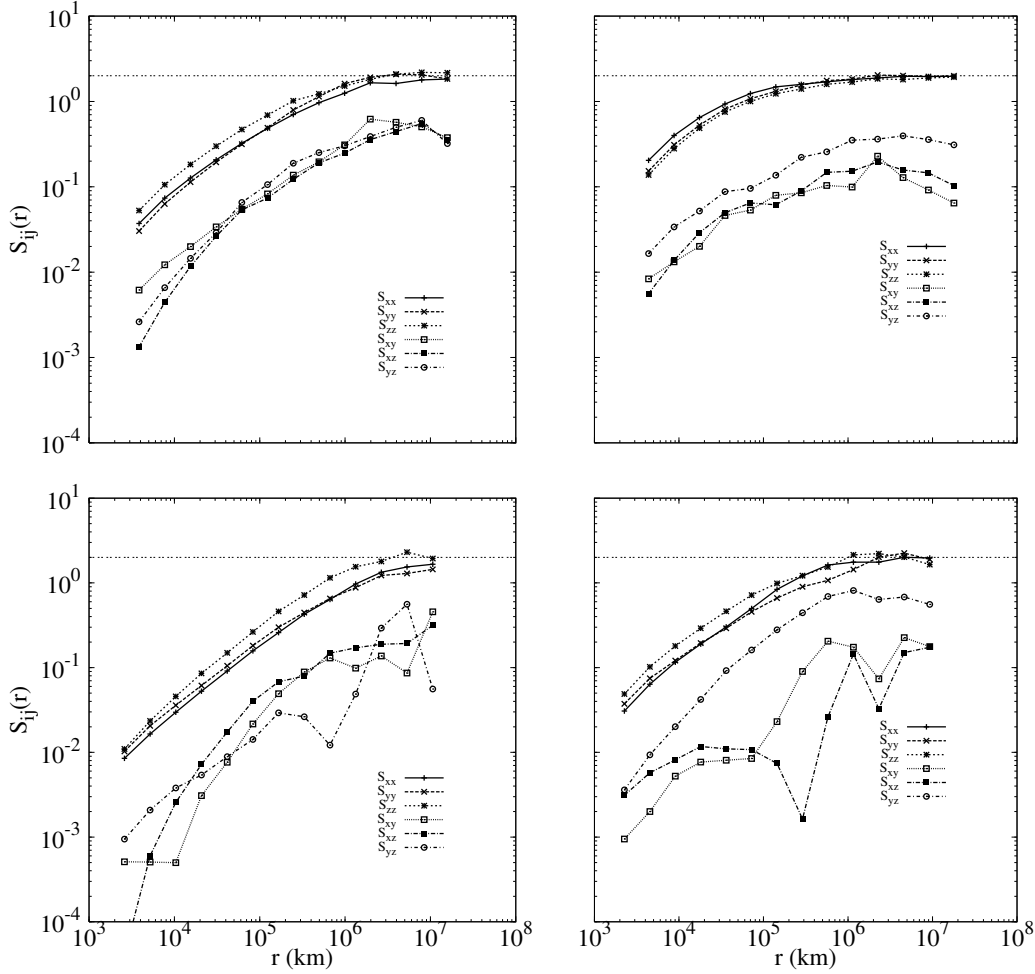}}
\caption{The second order structure function $S_{\alpha\beta}(r)$ of interplanetary magnetic field within fast (top panels) and slow (bottom panels) wind, both at $0.9$ AU (left panels) and $0.3$ AU (right panels), plotted versus the scale in log-log axes. The structure functions were computed using for each scale the fluctuations of magnetic field normalised to their standard deviations, so that for a large scale uncorrelated field the value of second order structure functions should be $2$.} 
\label{fig1}
\end{figure}

As a first analysis, let us discuss the scaling of the normalized longitudinal, transverse, and off-diagonal components of the tensor mentioned above. Figure~\ref{fig1} displays the scaling of the second order longitudinal, transverse, and off-diagonal components, for both fast and slow wind, at $0.3$AU and $0.9$AU. Each variable has been normalized to its own standard deviation before computing the structure functions.

Even at a first glance, it seems evident that the diagonal elements are more or less of the same order at all scales. In fact, it is possible to note a scaling from the large scale values ($S_{\alpha\alpha}=2$ for uncorrelated fields) toward zero, for almost all datasets, with at most a ratio of order $3$ for all scales. Those differences are more evident far from the sun for fast wind, while for slow wind they are already present at $0.3$AU. 
The off-diagonal elements are about one order of magnitude smaller than the diagonal elements. For isotropic fields, they should vanish at large scale because of decorrelation. There are indications that this is indeed the case, as can be seen looking at the rightmost points of the plots in Figure~\ref{fig1}, and taking into account the ``large scale'' values reported in Table~\ref{table_scales} (see below). However, most of the mixed components show the same scaling as the diagonal elements, that is they do not seem to vanish at small scales faster than the diagonal elements. In particular, in most of the cases the scaling at small scales is more or less the same for off-diagonal and for diagonal elements, so that the anisotropic contribution could eventually survive at small scale. This can be seen, for example, by looking at fast wind close to the earth, where $S_{xy}$ decreases as slowly as the diagonal $S_{\alpha\alpha}=2$, while $S_{xz}$ decreases more rapidly. This important result indicates that isotropy, in fact, is not restored at small scales. 
In slow wind, the off-diagonal elements seem to be more scattered and, in general, slightly higher than in fast wind. This would mean that fast wind remains more anisotropic than slow wind. 
Concerning radial evolution, it might be said that at short distance from the sun the anisotropic contributions (off-diagonal terms) are less important, while they increase approaching the Earth. These observations suggest that perhaps the anisotropy may be hidden by the Alfv\'enic nature of turbulence, which is more evident in fast wind and near the sun (see for example Ref.s \cite{RevBC,bruno_rad}), while the angle between large scale magnetic field (the Parker spiral profile) and mean flow is of course more relevant as it is increased, far from the sun. 
This behaviour is even more evident by looking at the fourth order moments, while for the odd orders moments the scaling is unclear and no information can be inferred (not shown)~\cite{relief}.

At variance with Ref.~\cite{bigazzi}, since $S_{\alpha_1,\alpha_2,\cdots,\alpha_n}(r)$ display multiple power-law scaling, we prefer to fit the structure functions by the following Batchelor relation~\cite{majda,kurien1,kurien2}

   \begin{equation}
S_{\alpha_1,\cdots,\alpha_n}(r) = \frac{A_{\alpha_1,\cdots,\alpha_n}\eta^n \left(r/\eta\right)^n}{\left[1+B_{\alpha_1,\cdots,\alpha_n} \left(r/\eta\right)^2\right]^{C_{\alpha_1,\cdots,\alpha_n}}}
\left[1+D_{\alpha_1,\cdots,\alpha_n}\left(r/L_0\right) \right]^{2C_{\alpha_1,\cdots,\alpha_n}-n}
  \label{funzione}
   \end{equation}
The free parameters in equation~(\ref{funzione}) include the main scales of the flow, which need to be estimated independently. In particular, $\eta$ is the dissipation scale, which can be estimated, for each stream, as the IMF typical cyclotron scale, and computed directly from the data as $\eta=2\pi m_p V_{th}/eB_0$, $m_p$ being the proton mass, $V_{th}$ the thermal speed and $e$ the electron charge. Note that this scale represents in fact lower bound for the actual dissipation scale, which cannot be directly measured from the data because of resolution limits. The integral scale $L_0$ is also obtained, for each stream, from the isotropic magnetic power spectra $E(k)$ as $L_0=\int{E(k)k^{-1}dk}/\int{E(k)dk}$. Values of $L_0$ and $\eta$ are reported in Table~\ref{table_scales}. The integral scale $L_0$ increases with the distance for both types of wind, and is larger for slow wind. This seems to indicate that turbulence is more developed in slow wind. Since the integral scale is growing moving away from the sun \cite{RevMT,RevBC}, but at the same time the estimated dissipation scale increases more rapidly, the inertial range and thus the Reynolds number get smaller moving away from the sun, as happens in decaying turbulence~\cite{frisch}, and in disagreement with previous analysis~\cite{pagel}.

\begin{table}
\centering \caption{The typical scales of the flow $L_0$ and $\eta$ (see text),
for fast (F) and slow (S) wind, at $R=0.3$AU (@0.3) and $R=0.9$AU (@0.9).}
\label{table_scales}
\begin{tabular}{ccc}
\hline\noalign{\smallskip} \centering
 Stream & $L_0$ ($10^6$Km) & $\eta$ (Km)  \\
\noalign{\smallskip}\hline\noalign{\smallskip}
Fast Wind $R = 0.3$& $5.0$ & $578$ \\
Fast Wind  $R = 0.9$& $11.9$ & $1920$ \\
Slow Wind $R = 0.3$& $14$ & $150$ \\
Slow Wind $R = 0.9$& $16.3$ & $2520$ \\
\noalign{\smallskip}\hline
\end{tabular}
\end{table}

\begin{figure}
\epsfxsize=8cm 
\centerline{\epsffile{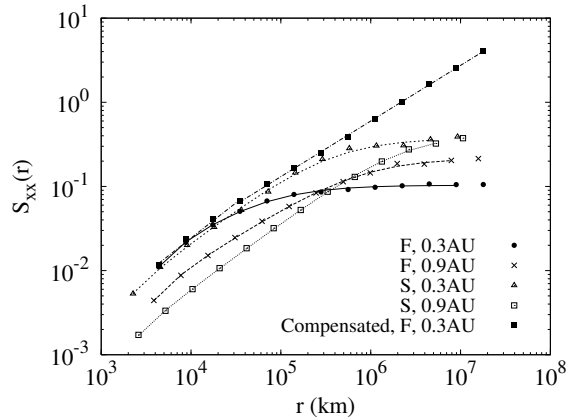}} 
\caption{The second-order longitudinal structure function $S_{xx}(r)$ of interplanetary magnetic field within fast and slow wind at $0.9$ and $0.3$ AU, plotted versus the scale $r$ in log-log axes, together with one example of the compensated function (see legend). Lines represent the fits with the function~(\ref{funzione}), while the compensated function has been fitted with a power-law with exponent $0.67\pm 0.02$.} 
\label{fig2}
\end{figure}

The multiple power-law behavior of the structure functions is described by equation~(\ref{funzione}) through the exponent $C_{\alpha_1,\alpha_2,\cdots,\alpha_n}$, which are the scaling parameters obtained from the fit of the data corresponding to the inertial range scaling, such that the usual scaling exponent can be recovered as $\zeta_{\alpha_1,\cdots,\alpha_n} = n-2C_{\alpha_1,\cdots,\alpha_n}$~\cite{kurien1,kurien2}. This can easily be tested by dividing the structure functions by the factor $\left[1+D_{\alpha_1,\cdots,\alpha_n}\left(r/L_0\right)\right]^{2C_{\alpha_1,\cdots,\alpha_n}-n}$, and observing the power-law behavior, with exponent $\zeta_{\alpha_1,\cdots,\alpha_n}$, of the resulting extended structure function. 
Figure~\ref{fig2} shows the behaviour of some second order mixed structure functions, for both fast and slow wind, and at two heliocentric distances, together with the fit with equation~(\ref{funzione}), as well as the compensated function and a power-law fit for one case. It is evident that equation (\ref{funzione}) can reproduce the data with very good agreement, so that the values obtained for the reduced $\chi^2$ are always less than unity. The most striking differences between structure functions computed for fast and slow wind are shown by the data samples recorded close to the sun ($0.3$ AU). In addition, a remarkable radial evolution is clearly visible, and it is stronger for fast wind, as already observed in different contexts \cite{RevMT,RevBC}. 

The values of the exponents obtained from the  fit of the second and fourth order structure functions tensor longitudinal and fully anisotropic components with equation~(\ref{funzione}) are listed in table \ref{table1}, for both fast and slow wind, at $0.3$ AU and $0.9$ AU. For both slow and fast wind the differences between the longitudinal and the anisotropic components are small. Once more this implies that the anisotropic contribution does not vanish at small scales, but its contribution to the scaling laws is, at least, decreasing as slowly as the isotropic part. 

\begin{table*}
\begin{center}
\caption{Scaling exponents $\zeta_{\alpha_1\cdots\alpha_n}$ of the longitudinal and fully anisotropic components of the second (marked with $*$) and fourth order structure function components for fast (left columns) and slow (right columns) solar wind streams, at two different distances $R$ from the Sun.} \label{table1}
\begin{tabular}{ccccc}
\hline\noalign{\smallskip} \centering
 & FAST WIND& & $\; \;$ SLOW WIND& \\
\noalign{\smallskip}\hline\noalign{\smallskip}
$\zeta_{\alpha_1\cdots\alpha_n}\; \;$ & $R = 0.3$ AU & $R = 0.9$ AU  & $R = 0.3$ AU & $R = 0.9$ AU \\
\noalign{\smallskip}\hline\noalign{\smallskip}
$\zeta_{xx}$ & $0.67 \pm 0.02$ & $0.66 \pm 0.03$ & $0.60 \pm 0.03$ & $0.60 \pm 0.02$ \\
$\zeta_{yy}$ & $0.69 \pm 0.1$ & $0.62 \pm 0.03$ & $0.68 \pm 0.03$ & $0.62 \pm 0.02$ \\
$\zeta_{zz}$ & $0.70 \pm 0.1$ & $0.66 \pm 0.04$ & $0.68 \pm 0.03$ & $0.57 \pm 0.02$ \\
$\zeta_{xy} (*)$ & $0.68 \pm 0.1$ & $0.66 \pm 0.03$ & $0.64 \pm 0.03$ & $0.60 \pm 0.02$ \\
$\zeta_{xz} (*)$ & $0.53 \pm 0.02$ & $0.67 \pm 0.03$ & $0.57 \pm 0.02$ & $0.59 \pm 0.02$ \\
$\zeta_{yz} (*)$ & $0.71 \pm 0.09$ & $0.59 \pm 0.02$ & $0.67 \pm 0.03$ & $0.62 \pm 0.02$ \\
\hline
$\zeta_{xxxx}$ & $1.68 \pm 0.01$ & $1.19 \pm 0.03$ & $1.31 \pm 0.01$ & $1.18 \pm 0.04$ \\
$\zeta_{xzzz} (*)$ & $1.51 \pm 0.1$  & $1.54 \pm 0.03$ & $1.43 \pm 0.04$ & $1.33 \pm 0.03$ \\
$\zeta_{xxxz} (*)$ & $1.52 \pm 0.1$  & $1.55 \pm 0.03$ & $1.39 \pm 0.04$ & $1.37 \pm 0.02$ \\
\hline
\noalign{\smallskip}\hline
\end{tabular}
\end{center}
\end{table*}

In conclusion, we have investigated anisotropy of magnetic fluctuations as they are measured in the interplanetary space, by using a method which allows to extract anisotropic exponents by avoiding mixing with the isotropic sector~\cite{kurien1,kurien2,bigazzi}. The normalized isotropic longitudinal components, and the fully anisotropic components of the second and higher orders structure function tensor are shown to have similar scaling toward the small scales, where anisotropy effects could be expected to vanish. This can be interpreted as an evidence that small scale anisotropy is present. Then, following Kurien and Sreenivasan \cite{kurien1,kurien2} scaling exponents have been extracted by using a scaling function that interpolates the high-order tensor components. The comparison between the scaling exponents of the longitudinal and of the anisotropic components shows once again that small-scale fluctuations remain anisotropic. Thus, the return-to-isotropy invoked in the usual fluid flows does not work in MHD turbulence to a very large extent.

\acknowledgments{We are grateful to Alain Noullez for useful discussion and to an anonymous referee for helpful comments.}

\end{document}